\documentclass[twocolumn,preprintnumbers,amsmath,amssymb,floatfix,superscriptaddress]{revtex4-1}
\usepackage{amsmath,hyperref,amssymb}
\usepackage{graphicx}
\usepackage{dcolumn}
\usepackage{bm}
\usepackage{verbatim}
\usepackage{tabularx}
\usepackage{slashed}
\usepackage{cancel}
\usepackage{hyperref}
\usepackage[utf8]{inputenc}
\usepackage{feynmf}
\usepackage[usenames,dvipsnames]{color}

\def\be{\begin{equation}}
\def\ee{\end{equation}}
\def\bea{\begin{eqnarray}}
\def\eea{\end{eqnarray}}
\def\ba#1\ea{\begin{align}#1\end{align}}
\def\bg#1\eg{\begin{gather}#1\end{gather}}
\def\bm#1\em{\begin{multline}#1\end{multline}}
\def\bmd#1\emd{\begin{multlined}#1\end{multlined}}
\newcommand{\eref}[1]{(\ref{#1})}

\begin{document}

\title{Non-Abelian bosonization and modular transformation approach to superuniversality}

\author{Aaron Hui}
\affiliation{\small \it School of Applied \& Engineering Physics, Cornell University, Ithaca, New York 14853, USA}
\author{Eun-Ah Kim}
\affiliation{\small \it Department of Physics, Cornell University, Ithaca, NY, 14853, USA}
\author{Michael Mulligan}
\affiliation{\small \it Department of Physics and Astronomy, University of California,
Riverside, CA 92511, USA}

\date{\today}

\begin{abstract}
Quantum Hall inter-plateaux transitions are physical exemplars of quantum phase transitions.
Near each of these transitions, the measured electrical conductivity scales with the same correlation length and dynamical critical exponents, i.e., the critical points are superuniversal.
In apparent contradiction to these experiments, prior theoretical studies of quantum Hall phase transitions within the framework of Abelian Chern-Simons theory coupled to matter found correlation length exponents that depend on the value of the quantum critical Hall conductivity.
Here, we use non-Abelian bosonization and modular transformations to theoretically study the phenomenon of superuniversality.
Specifically, we introduce a new effective theory that has an emergent $U(N)$ gauge symmetry with any $N > 1$ for a quantum phase transition between an integer quantum Hall state and an insulator. 
We then use modular transformations to generate from this theory effective descriptions for transitions between a large class of fractional quantum Hall states whose quasiparticle excitations have Abelian statistics. 
We find the correlation length and dynamical critical exponents are independent of the particular transition within a controlled 't Hooft large $N$ expansion, i.e., superuniversal!
We argue that this superuniversality could survive away from this controlled large $N$ limit using recent duality conjectures.
\end{abstract}

\maketitle

\section{Introduction}

As a two-dimensional electron gas is tuned by a perpendicular magnetic field from one quantum Hall state to another, the longitudinal electrical resistance exhibits a peak with a width $\Delta B \propto T^{1/\nu z}$, where $\nu$ and $z$ are correlation length and dynamical critical exponents and $T$ is the temperature; the slope of the Hall resistance likewise diverges as $\Delta B$ as a particular transition is approached \footnote{Similar scaling is observed if the temperature is replaced by frequency, applied current, or inverse system size.}.
The surprising feature is that the observed $\nu \approx 7/3$ and $z \approx 1$ appear to be insensitive to whether the transition is between integer or fractional Abelian quantum Hall states \cite{PhysRevLett.61.1294, ENGEL199013, PhysRevLett.67.883, PhysRevLett.71.2638, PhysRevB.50.14609, PhysRevB.51.18033, PhysRevLett.94.206807, PhysRevLett.102.216801, PhysRevB.81.033305} (See note \footnote{Strictly speaking, the product $\nu z$ has only been factorized at integer plateau transitions, however, dimensional analysis suggests $z = 1$ for both types of transitions if the Coulomb interaction provides the dominant scale.})
Taken at face value, the implication is that the associated quantum critical points \cite{SondhiGirvinCariniShahar, SachdevQPT} have the same critical indices for comparable observables \cite{PhysRevB.32.1311, PhysRevLett.64.1297, Kivelson1992, Lutken:1991jk, Fradkin:1996xb, ShimshoniSondhiShahar1997, PhysRevB.63.155309, Goldman:2018zfm} and are instead distinguished by their critical conductivity \cite{PhysRevLett.70.481, Shahar1995, PhysRevLett.77.4426} (see \footnote{See \cite{PhysRevLett.70.3796} for an explanation of the low-temperature conductivity in the scaling region in terms of variable-range hopping.}); 
this phenomenon is known as {\it superuniversality} \cite{ShimshoniSondhiShahar1997}.

The root cause of superuniversality has been a puzzle since its observation over three decades ago.
Numerical studies of the integer quantum Hall transition, modeled by disordered noninteracting electrons, find a correlation length exponent in qualitative agreement with experiment \cite{chalkercoddington, RevModPhys.67.357, PhysRevB.80.041304}, however, these theories have $z \approx 2$ and it is challenging to generalize these works to transitions between fractional quantum Hall states \cite{PhysRevLett.70.4130}.
Theories of Abelian Chern-Simons gauge fields coupled to matter, i.e., theories of composite bosons or composite fermions \cite{PhysRevLett.58.1252, jain1989, Read89, zhang1989, lopezfradkin91, halperinleeread, Kalmeyer1992}, provide a unifying, physical framework for studying both integer and fractional quantum Hall transitions.
Thus far, controlled approximations to these strongly coupled theories, obtained when the number of fermion or boson flavors is large and there is no disorder, have failed to yield superuniversal behavior: the calculated correlation length exponent depends on the particular quantum Hall transition \cite{wenwu1993, ChenFisherWu1993, 2016PhRvB..93c5103L, PhysRevB.97.085112}.
It is important to determine whether these calculations reveal a generic behavior of the field theoretical models or, instead, reflect certain artifacts of the approximation scheme \footnote{Superuniversality has been found in studies of disordered Dirac fermions of various symmetry classes in $3+1$ dimensions \cite{2017PhRvB..95g5131G} and in certain models with long-ranged statistical interactions \cite{2012geraedtsmotrunich} $2+1$ dimensions.}.
In this paper, we provide evidence for the latter.

\begin{figure}[h!]
\centering
\includegraphics[width= .85 \linewidth]{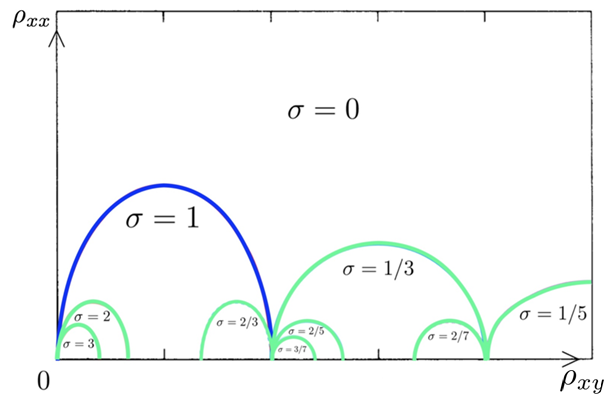}
\caption{Schematic zero-temperature phase diagram \cite{Kivelson1992} in the space of Hall $\rho_{xy}$ and longitudinal resistivity $\rho_{xx}$.
Phases are denoted by their zero-temperature complex conductivity $\sigma = \sigma_{xy} + i \sigma_{xx}$, measured in units of $e^2/h$. The blue boundary denotes the $1\rightarrow 0$ integer quantum Hall transition, while the green boundaries denote transitions we derive from the $1\rightarrow 0$ transition via modular transformations.
}
\label{phasediagram}
\end{figure}
As a step towards understanding the observed behavior, we focus here on the fundamental theoretical question raised by the appearance of superuniversality, i.e., how distinct interacting critical points can share the same critical exponents. 
To this end, we introduce new theories, involving a single Dirac fermion coupled to a {\it non-Abelian} $U(N)$ Chern-Simons gauge field for any $N>1$, that exhibit quantum phase transitions between Abelian quantum Hall states. 
Intuitively, the $U(N)$ gauge symmetry of our theories generalizes the Abelian gauge symmetry implementing flux attachment in familiar composite boson/fermion theories.
In fact, as demonstrated in Appendix \ref{dualityargument}, these $U(N)$ gauge theories are dual to theories with an Abelian group.
The advantage of the enlarged gauge group is that it motivates an alternate approximation to our strongly coupled theories, namely, a controlled 't Hooft large $N$ expansion \cite{tHooft:1973alw} 
 \footnote{See \cite{StanfordGroup2014, 2016arXiv160705725W, PhysRevB.92.205104, PhysRevLett.117.157001, 2016JHEP...04..103G} for earlier applications of the large $N$ expansion in condensed matter physics. 
Gauge/gravity duality can provide an alternative framework where large $N$ naturally appears \cite{2008JHEP...11..020D}.}, within which we find that superuniversality occurs without the inclusion of disorder.

We emphasize that the theories we consider here have more symmetries than the physical systems motivating our work; for instance, our theories are Lorentz-invariant and, in particular, preserve translational invariance.
Our hope is that our theories might represent ``parent'' theories for more realistic descriptions of the experimental systems.
Consequently, we defer quantitative questions specific to the particular experimental systems to the future.

The remainder of this paper is organized as follows.
In Sec.~\ref{integersection}, we introduce a new description for an integer quantum Hall transition; this theory is inspired by fermion particle-vortex duality \cite{Son2015, WangSenthilfirst2015, MetlitskiVishwanath2016, PhysRevX.6.031043,Seiberg:2016gmd} (see the related work \cite{Kachru:2015rma,Geraedtsetal2015,XuYou2015selfdual, MrossAliceaMotrunichexplicitderivation2016, MrossAliceaMotrunichbosonicph2016}) and various bosonization dualities in $2+1$ dimensions \cite{GMPTWY2012, AharonyGurAriYacoby2012, AharonyGurAriYacobysecond2012,JainMinwallaYokoyama2013, Gur-AriYacoby2015, Aharony2016, 2016arXiv160601912M,Hsin:2016blu,PhysRevD.94.085009,PhysRevLett.118.011602, 2017arXiv170505841C, 2017arXiv170501106M, 2017PhRvX...7c1051W}.
Guided by Ref.~\cite{Kivelson1992}, where the phase diagram in Fig.~\ref{phasediagram} was proposed by
extending the theory of two-parameter scaling of the Hall and longitudinal resistivity for the integer Hall effect \cite{Levine:1984yg} to the fractional Hall regime using the ``law of corresponding states" \cite{PhysRevLett.64.1297, Kivelson1992}, we then derive in Sec.~\ref{modularsection} new effective theories for various fractional quantum Hall transitions using modular transformations \cite{WittenSL2Z2003}.
In Sec.~\ref{controlledthooft}, we show that the correlation length and dynamical critical exponents of our theories are insensitive to the particular quantum Hall phase transition within a controlled 't Hooft large $N$ limit.
In Sec.~\ref{duality}, we discuss how recent duality conjectures imply that the physics of our $U(N)$ Chern-Simons coupled to matter theories is {\it independent} of $N > 1$.
This is the crucial feature that allows us to argue that critical exponents, calculated in the 't Hooft large $N$ limit, are exact at the leading planar order and that superuniversality may persist away from the controlled 't Hooft large $N$ limit. 
In addition, there are six appendices that discuss details of arguments presented in the main text.

\section{Integer quantum Hall transition} 
\label{integersection}

Our starting point is an effective Lagrangian for an integer quantum Hall transition,
\begin{widetext}
\begin{align}
\label{seed}
{\cal L}_{\rm IQHT}(A) & = i \bar{\psi} \not \! \! D_{a} \psi - M_\psi \bar{\psi} \psi - {1 \over 2} {1 \over 4 \pi} {\rm Tr} \Big[a d a - {2 \over  3} i a^3 \Big] 
- {1 \over 2 \pi} {\rm Tr}[a] d b - {N + 1 \over 4 \pi} b d b - {1 \over 2 \pi} b d A.
\end{align}
\end{widetext}
The notation is as follows:
$\psi$ is a two-component Dirac fermion transforming in the fundamental representation of $U(N)$;
$a$ and $b$ are dynamical $U(N)$ and $U(1)$ Chern-Simons gauge fields;
$A$ (above and throughout) is a non-dynamical Abelian gauge field that we think of as electromagnetism;
$\slashed{D}_a = \gamma^\mu (\partial_\mu - i a_\mu)$ for $\mu \in \{t,x,y\}$ and $\gamma$-matrices satisfying $\{\gamma^\mu, \gamma^\nu\} = 2 \eta^{\mu \nu}$ where $\eta^{\mu \nu} = {\rm diag}(1, -1, -1)$;
$\bar{\psi} = \psi^\dagger \gamma^t$;
$N$ is a positive integer; Abelian Chern-Simons terms $A d B = \epsilon^{\mu \nu \rho} A_\mu \partial_\nu B_\rho$, and the cubic interaction in the non-Abelian Chern-Simons term $a^3 = {1 \over 2}  \epsilon^{\mu \nu \rho} a_\mu a_\nu a_\rho$.
For simplicity of presentation, we regularize the theory in \eqref{seed} by implicitly including a Yang-Mills term for $a$ and Maxwell term for $b$ \cite{witten1989, Chen:1992ee}.
See Appendix \ref{chernsimonsdefs} 
for further explanation of the notation and for a few pertinent facts about Chern-Simons theories.

Prior work studying Chern-Simons gauge theories coupled to matter suggests that the theory in $\eqref{seed}$ realizes a critical point with conformal symmetry \cite{Chen:1992ee, Avdeev:1992jt}.
In Appendix \ref{dualityargument}, we argue nonperturbatively that this critical point is in the free Dirac fermion universality class for any integer $N \geq 1$.

For the moment, we verify that \eqref{seed} describes a transition between an integer quantum Hall state and an insulator as the fermion mass $M_\psi$ is tuned through zero, consistent with our identification in Appendix \ref{dualityargument} of \eqref{seed} with the theory of a free Dirac fermion. 
See Appendix \ref{integerphases} for additional details. 
Remarkably, this demonstration applies for any integer $N \geq 1$.
In our theory, the mass $M_\psi$ represents an {\it effective} control parameter for a particular quantum phase transition.
For definiteness, it may be helpful to think about $M_\psi$ in terms of the analogous tuning parameter that appears in lattice models for integer quantum Hall transitions \cite{PhysRevLett.61.2015, PhysRevB.50.7526}.
In these latter models, the transition is controlled by the ratio of the on-site chemical potential to the second nearest-neighbor hopping. This theory matches the realistic integer quantum Hall transition insofar that it describes some transition between two integer quantum Hall states, as is commonly done in the literature.

Our strategy is to identify the insulating and integer quantum Hall states through their electrical response to the electromagnetic gauge field $A$.
Below the energy scale of the mass $|M_\psi|$, we can integrate out $\psi$ to obtain:
\begin{align}
\label{effectiveinteger}
{\cal L}_{\rm eff} & = {{\rm sign}(M_\psi) - 1 \over 2} {1 \over 4 \pi} {\rm Tr} \Big[a d a - {2 \over  3} i a^3 \Big]  \cr
&  - {1 \over 2 \pi} {\rm Tr}[a] d b - {N + 1 \over 4 \pi} b d b - {1 \over 2 \pi} b d A.
\end{align}
In this effective Lagrangian, only relevant and marginal terms in the renormalization group sense are written.
If $M_\psi < 0$, rank/level duality \cite{NACULICH1990687, Nakanishi1992, Hsin:2016blu} (Appendix \ref{integerphases}) implies that
\begin{align}
\label{iqhe}
{\cal L}_{\rm eff}(M_\psi < 0) = {1 \over 4 \pi} A d A,
\end{align}
i.e., the effective electrical response Lagrangian of an integer quantum Hall state.
Consequently, we identify the phase obtained for $M_\psi < 0$ with an integer quantum Hall state.
Integrating out fermions with $M_\psi > 0$, the non-Abelian Chern-Simons term for $a$ disappears.
Only ${\rm Tr}[a]$ remains in the effective Lagrangian; the $SU(N) \subset U(N)$ component of $a$ decouples and we assume it confines \cite{FEYNMAN1981479}.
The equation of motion for ${\rm Tr}[a]$ sets $b = 0$ \cite{WittenSL2Z2003} and the resulting effective Lagrangian,
\begin{align}
\label{insulator}
{\cal L}_{\rm eff}(M_\psi > 0) = 0,
\end{align}
describes an electrical insulator.
We expect the leading irrelevant operator supplementing the effective Lagrangian in Eq.~\eqref{insulator} to be a Maxwell term for $A$, consistent with our identification of the phase obtained when $M_\psi > 0$ with an insulator.

\section{Generating fractional quantum Hall transitions}
\label{modularsection}

We now show how to generate effective descriptions with $U(N)$ gauge symmetry for fractional quantum Hall transitions using the modular group, $PSL(2,\mathbb{Z})$, i.e.,
the group of $2 \times 2$ matrices with integer entries and unit determinant.
On a complex number, like the complexified zero-temperature dc conductivity $\sigma = \sigma_{xy} + i \sigma_{xx}$ (measured in units of $e^2/h$) \footnote{We define $\sigma = \lim_{\omega \rightarrow 0} \lim_{T \rightarrow 0} \sigma(\omega, T)$.
We only require $\sigma$ for quantum Hall states described at long distances by Chern-Simons theory, since we only need to know how the Hall conductivity changes across a transition.
We caution that the order of limits may not generally commute for either gapped \cite{Zhang1992} or gapless states \cite{damlesachdev97}.}, the modular group takes 
\begin{align}
\label{modularconventional}
\sigma \mapsto {p \sigma + q \over r \sigma + s},\ \text{for}\ \begin{pmatrix} p & q \cr r & s \end{pmatrix} \in PSL(2,\mathbb{Z}).
\end{align}
Because the modular group is generated by two elements, ${\cal T} = \begin{pmatrix} 1 & 1 \cr 0 & 1 \end{pmatrix}$ and ${\cal S} = \begin{pmatrix} 0 & 1 \cr -1 & 0 \end{pmatrix}$, any element of $PSL(2,\mathbb{Z})$ can be decomposed into a product of ${\cal T}$ and ${\cal S}$ operators.

Ref.~\cite{WittenSL2Z2003} showed how the modular group in Eq.~\eqref{modularconventional} acts on the Lagrangian of a conformal field theory with $U(1)$ global symmetry.
(See \cite{LeighPetkouSL2Z2003} for the effects on higher-spin currents.)
Denoting the Lagrangian of an arbitrary conformal field theory by ${\cal L}(\Phi, A)$, where $\Phi$ collectively represents all dynamical fields and $A$ is a background field coupling to the $U(1)$ symmetry, the modular group acts as follows:
\begin{align}
\label{modulardef}
& {\cal T}: {\cal L}(\Phi, A) \mapsto {\cal L}(\Phi, A) + {1 \over 4 \pi} A d A, \cr
& {\cal S}: {\cal L}(\Phi, A) \mapsto {\cal L}(\Phi, c) - {1 \over 2 \pi} c d B.
\end{align}
Eq.~\eqref{modulardef} induces the action of the modular group on the complexified conductivity of the $U(1)$ symmetry of ${\cal L}(\Phi, A)$.
${\cal T}$ simply shifts the Hall conductivity by one unit;
${\cal S}$ inverts $\sigma \rightarrow - 1/\sigma$ through its replacement of $A$ with a dynamical $U(1)$ gauge field $c$ and introduction of a new background field $B$ via the coupling $-{1 \over 2\pi} c d B$.

Reminiscent of the ``law of corresponding states" \cite{Kivelson1992} (see Fig.~\ref{phasediagram})  we can generate using Eq.~\eqref{modulardef} an effective description for a transition between any two quantum Hall states related by a modular transformation to either the insulator ($\sigma = 0$) or integer quantum Hall state ($\sigma = 1$).
The pertinent subset of transformations can be decomposed into two operations: 
\begin{enumerate}
\item[(i)] addition of a Landau level = ${\cal T}$;
\item[(ii)] attachment of $m$ units of flux = ${\cal S}^{-1} {\cal T}^{- m} {\cal S}$. 
\end{enumerate}
Any transition from $\sigma = j \rightarrow j - 1$ between integer quantum Hall states is found by adding $j-1$ Landau levels, i.e., applying ${\cal T}^{j - 1}$ with $j \in \mathbb{Z}$ to Eq.~\eqref{seed}.
On the other hand, the fractional quantum Hall transition, $1/(m+1) \rightarrow 0$, is obtained by applying ${\cal S}^{-1} {\cal T}^{- m} {\cal S}$ to Eq.~\eqref{seed}.
We can combine the operations of adding a Landau level and flux attachment to find a description for the $1/3 \rightarrow 2/5$ transition using ${\cal S}^{-1} {\cal T}^{-2} {\cal S} {\cal T}$.
The $2/3 \rightarrow 1$ transition -- the lowest Landau level particle-hole conjugate of the $1/3 \rightarrow 0$ transition -- is obtained by acting on the Lagrangian in Eq.~\eqref{seed} with ${\cal T} {\cal S}^{-1} {\cal T}^{2} {\cal S} {\cal T}^{-1}$.
Other transitions can be generated by further iteration of these methods. Hence, modular transformations formalize the ``law of corresponding states" \cite{jain1989, PhysRevLett.64.1297, Kivelson1992}. 
Because we have not included effects of disorder, we are, in a sense, effectively considering the horizontal axis of Fig.~\ref{phasediagram}.

In the remainder of the paper, we focus on the ${1 \over m+1}\rightarrow 0$ transition where the even integer $m \geq 0$; qualitatively similar arguments apply for other transitions.
Applying the modular transformation described above to \eqref{seed}, we find the Lagrangian,
\begin{align}
\label{generallag}
{\cal L}_m & = {\cal L}_{\rm IQHT}(c) + {\cal L}_{\rm mod}(A),
\end{align}
where ${\cal L}_{\rm IQHT}(c)$ is given in Eq.~\eref{seed} with the replacement $A \rightarrow c$ and
\begin{align}
\label{simplefraction}
{\cal L}_{\rm mod}(A) = - {1 \over 2 \pi} c d g - {m \over 4 \pi} g d g + {1 \over 2 \pi} g d A,
\end{align} 
with dynamical $U(1)$ gauge fields $c$ and $g$.
Thus, the modular transformation simply introduces additional Chern-Simons gauge fields coupling to the $U(1) \subset U(N)$  gauge field ${\rm Tr}[a]$ in ${\cal L}_{\rm IQHT}$.
Appendix \ref{effectiveU1level} lists the corresponding effective Lagrangians, analogous to Eqs.~\eqref{generallag} and \eqref{simplefraction}, for other simple quantum Hall transitions.
When $m=0$, we may integrate out $c$ and $g$ using their equations of motion to recover the Lagrangian in Eq.~\eqref{seed}; when $m \geq 2$, we can no longer integrate out $g$ to recover an effective Lagrangian whose Chern-Simons terms have integer levels. 

It is straightforward to check (see Appendix \ref{appendixB} for details) using the arguments given below Eq.~\eqref{effectiveinteger} that ${\cal L}_m$ in Eq.~\ref{generallag} and its generalizations describe a large class of fractional quantum Hall phase transitions, tuned by the fermion mass.
We assume these transitions are continuous for any $m \geq 0$.

\section{Superuniversality in the 't Hooft large $N$ limit}
\label{controlledthooft}

Our goal is to determine the correlation length and dynamical critical exponents exhibited by ${\cal L}_m$ in Eq.~\eqref{generallag} for $m \geq 0$.
The (inverse) correlation length exponent, $\nu^{-1} = 1 - \gamma_{\bar{\psi} \psi}$, measures the anomalous dimension $\gamma_{\bar{\psi} \psi}$ of the operator $\bar{\psi} \psi(x)$ \footnote{The anomalous dimension is determined by the decay of the two-point function $\protect\langle \bar{\psi} \psi(x) \bar{\psi} \psi(0) \protect\rangle \sim |x|^{-2(1 + \gamma_{\bar{\psi} \psi})}$}, whose coefficient $M_\psi$ is the tuning parameter for the various fractional quantum Hall transitions we consider.
Since our effective theories are Lorentz-invariant, $z=1$ automatically.
Because ${\cal L}_m$ depends on the rank $N$ of the $U(N)$ gauge group of $a$, it is necessary to choose a particular value of $N$ at which to evaluate $\nu$.
We choose $N=\infty$ and determine $\nu$ in a controlled 't Hooft large $N$ limit.
In Sec.~\ref{duality}, we will argue that the physics of ${\cal L}_m$ is independent of $N$.
Consequently, $N=\infty$ represents a reliable value of the parameter $N$ at which to determine the critical exponents of ${\cal L}_m$.

In order to determine the correlation length exponent, it is helpful to first simplify the Lagrangian ${\cal L}_m$ as follows: we set the background field $A = 0$;
next, we integrate out all Abelian gauge fields (i.e., $b$, $c$, and $g$) not minimally coupled to $\psi$;  
finally, we decompose $a = a_{SU(N)} + a_{U(1)} \mathbb{I}$, where $a_{SU(N)}$ is a $SU(N) \subset U(N)$ gauge field, $a_{U(1)}$ is an Abelian gauge field, and $\mathbb{I}$ is the $N \times N$ identity matrix.
After performing these steps, ${\cal L}_m$ becomes
\begin{align}
\label{simpleexplicit}
{\cal L}_{\rm s} & = i \bar{\psi} \not \! \! D_{a} \psi + {k_{U(1)} \over 4 \pi} a_{U(1)} d a_{U(1)} \cr
& + {k_{SU(N)} \over 4 \pi} {\rm Tr} \Big[a_{SU(N)} d a_{SU(N)} - {2 \over  3} i a_{SU(N)}^3 \Big],
\end{align}
with $\displaystyle k_{U(1)}= {N^2 - N - N m \over 2(N+1+m)}\ \text{and}\ k_{SU(N)} = - {1 \over 2} - N$.
We included the one-loop exact correction \cite{witten1989, Chen:1992ee} 
to the $SU(N) \subset U(N)$ Chern-Simons level $k_{SU(N)}$.
Although ${\cal L}_{\rm s}$ obscures the topological structure of our quantum critical state and any gapped phase obtained from it when $M_\psi \neq 0$ \footnote{The allowed Wilson loop observables are not manifest in Chern-Simons Lagrangians with non-quantized levels.
If the Chern-Simons theory is to describe a gapped state, additional information is needed to specify the quasiparticle spectrum.}, the perturbative analysis is unchanged.

To gain some intuition for the possible behavior of ${\cal L}_{\rm s}$ (and, therefore, ${\cal L}_m$), suppose the fluctuations of $a_{SU(N)}$ were ignored.
Then, ${\cal L}_{\rm s}$ would effectively describe $N$ flavors of fermions interacting with the Abelian Chern-Simons gauge field $a_{U(1)}$.
For such theories, it is known that $\gamma_{\bar{\psi} \psi} = 1 + {\cal O}({1 \over k_{U(1)} N})$ at large $N$ \cite{ChenFisherWu1993}.
Because $k_{U(1)} \propto N$ as $N \rightarrow \infty$ for any fixed $m$, the effects mediated by $a_{U(1)}$ could then be made arbitrarily small as $N \rightarrow \infty$.
(This is true for the other quantum Hall transitions considered in Appendix \ref{effectiveU1level}.)
Consequently, since $m$ only appears in $k_{U(1)}$, $\gamma_{\bar{\psi} \psi}$ would be independent of $m$ at $N=\infty$, i.e., superuniversal.
Our task now is to determine the extent to which this conclusion survives the inclusion of $a_{SU(N)}$ fluctuations.
 
The 't Hooft large $N$ limit \cite{tHooft:1973alw} (see \cite{Coleman} for a review) provides an expansion within which to calculate $\gamma_{\bar{\psi} \psi}$.
This limit, which is distinct from the limit that obtains within large flavor expansions, is defined by taking the rank of the $U(N)$ gauge group $N \rightarrow \infty$ with the ratios $\lambda_{SU(N)} = N/k_{SU(N)}$ and $\lambda_{U(1)} = N/k_{U(1)}$ held fixed.
Observables like $\gamma_{\bar{\psi} \psi}$ are then calculated in an expansion in powers of $1/N$; the coefficient of a particular power of $1/N$ is generally a power series in $\lambda_{SU(N)}$ and $\lambda_{U(1)}$.
In addition, there could be non-perturbative $\lambda_{SU(N)}$ and $\lambda_{U(1)}$ contributions to $\gamma_{\bar{\psi} \psi}$.
Our result in this section ignores any such non-perturbative corrections; our duality argument in the next section indicates such corrections are absent at least when $m=0$.

\begin{figure}[h!]
\centering
\includegraphics[width=1 \linewidth]{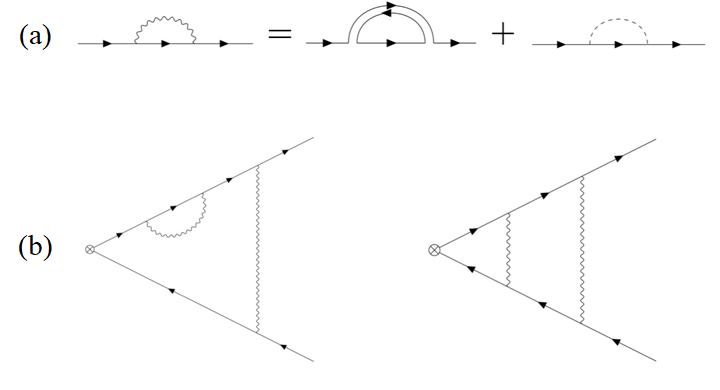}\\
\caption{(a) One-loop fermion self-energy decomposed into $SU(N) \subset U(N)$ and $U(1) \subset U(N)$ contributions. 
The closed oriented loop produces the relative factor of $N$ between the second and third diagrams.~(b) The leading Feynman diagrams contributing to $\gamma_{\bar{\psi} \psi}$ in the 't Hooft large $N$ limit.
Directed lines are fermion propagators; wavy lines are $U(N)$ gauge field propagators; a double line is a $SU(N)$ gauge field propagator; a dashed line denotes a $U(1)$ gauge field propagator; insertion of $\bar{\psi} \psi$ is represented by $\otimes$.}
\label{2loopdiagrams}
\end{figure}

As an illustrative example of how large $N$ scaling works, Fig.~\ref{2loopdiagrams}(a) decomposes the $a_{SU(N)}$ and $a_{U(1)}$ one-loop contributions to the fermion self-energy.
In our conventions, vertices scale as $N^0$, while gauge field propagators come with factors of $k_{SU(N)}^{-1}$ or $k_{U(1)}^{-1}$ depending upon whether $a_{SU(N)}$ or $a_{U(1)}$ propagates; $\psi$ propagators scale as $N^0$.
At large $N$, the $a_{SU(N)}$ contribution in Fig.~\ref{2loopdiagrams}(a) scales as $\lambda_{SU(N)}$, while the $a_{U(1)}$ correction scales as $\lambda_{U(1)}/N$.
(Here, we have assumed the $U(N)$ coupling constant achieves its fixed point value, proportional to $N^0$.)
Thus, the contribution of $a_{U(1)}$ in Fig.~\ref{2loopdiagrams}(a) is subdominant to that of $a_{SU(N)}$ as $N \rightarrow \infty$ by a factor of $1/N$.
This is a general feature: in perturbation theory, the 't Hooft large $N$ limits of $SU(N)$ and $U(N)$ gauge theories give identical results for shared observables \cite{Coleman}.
For Chern-Simons gauge theories with $U(N)$ gauge group, this relies on the $1/N$ suppression of diagrams containing closed loops of $a_{U(1)}$ relative to the corresponding planar diagrams that instead contain loops of $a_{SU(N)}$.

So long as $|k_{U(1)}| \propto N$ as $N \rightarrow \infty$, the effects of $a_{U(1)}$ are subdominant by a factor of $1/N$ in the 't Hooft large $N$ limit.
In particular, only the fluctuations of $a_{SU(N)}$ contribute to $\gamma_{\bar{\psi} \psi}$ at $N=\infty$.
The planar contribution to $\gamma_{\bar{\psi} \psi}$ scales with $N$ as $N^0$ and consists of an infinite expansion in $\lambda_{SU(N)}$; the first sub-planar contribution scales as $1/N$ and consists of an infinite series in $\lambda_{SU(N)}$ and $\lambda_{U(1)}$.
Thus, the 't Hooft expansion for $\gamma_{\bar{\psi} \psi}$ has the form:
\begin{align}
\label{expansion}
\gamma_{\bar{\psi} \psi} = f_0(\lambda_{SU(N)}) + {1 \over N} f_1(\lambda_{SU(N)}, \lambda_{U(1)}) + \ldots,
\end{align}
where the planar term $f_0(\lambda_{SU(N)})$ is a power series in $\lambda_{SU(N)}$, the first sub-planar term $f_1(\lambda_{SU(N)}, \lambda_{U(1)})$ is a power series in $\lambda_{SU(N)}$ and $\lambda_{U(1)}$, and $\ldots$ represent higher powers of $1/N$ which are expected to be subdominant in this expansion.
(The assumption that $f_0$ and $f_1$ are power series of their arguments is the statement that we are ignoring possible non-perturbative contributions to $\gamma_{\bar{\psi} \psi}$.)
Because $m$ only appears in $\lambda_{U(1)}$, through its appearance in the effective Chern-Simons level $k_{U(1)}$ of $a_{U(1)}$ (see Eq.~\eqref{simpleexplicit} and Appendix \ref{effectiveU1level}), $\nu$ is insensitive to the particular $1/(m+1) \rightarrow 0$ transition at $N = \infty$.
This is superuniversality in the 't Hooft large $N$ limit.

The specific value of $\nu$ is determined by $f_0(\lambda_{SU(N)})$ at $N=\infty$.
An important point is that the 't Hooft large $N$ limits of the theories we consider remain non-trivial even when $N = \infty$.
For instance, $|\lambda_{SU(N)}| = 1$ for $N = \infty$, so that an infinite number of terms generally need to be summed to determine $f_0(\lambda_{SU(N)})$.
Here, we find $\gamma_{\bar{\psi} \psi}$ in a {\it controlled} 't Hooft large $N$ limit, where it is necessary to continue $k_{SU(N)}$ away from its physical value (given below Eq.~\eqref{simpleexplicit}) such that $\lambda_{SU(N)} \ll 1$ and $f_0(\lambda_{SU(N)})$ can be reliably approximated by the leading terms in its expansion in $\lambda_{SU(N)}$. 

Figure~\ref{2loopdiagrams}(b) displays the leading contributions to $\gamma_{\bar{\psi} \psi}$ arising from the fluctuations of $a_{SU(N)}$ \footnote{The one-loop vertex diagram, as well as one-loop and two-loop fermion self-energy diagrams do not contain logarithmic divergences and, consequently, do not contribute to $\gamma_{\bar{\psi} \psi}$ \cite{Chen:1992ee, Avdeev:1992jt, GMPTWY2012}.}.
In \cite{GMPTWY2012}, it was shown that these two contributions cancel, i.e., $\gamma_{\bar{\psi} \psi} = 0$ to two-loop planar order or, equivalently, $f_0(\lambda_{SU(N)}) = 0$ to ${\cal O}(\lambda_{SU(N)}^2)$.
Thus, at the critical point described by ${\cal L}_m$ in  Eq.~\eqref{generallag}:
\begin{align}
\label{2loopcorrelationexponent}
\nu = 1 + {\cal O}\Big(\lambda_{SU(N)}^3 \Big),
\end{align}
for any $m \geq 0$ in the controlled 't Hooft large $N$ limit.
In perturbation theory, the dependence on $m$, i.e., the particular fractional quantum Hall critical point, appears at sub-planar order and is unobservable at $N = \infty$.

\section{$N$ independence and duality}
\label{duality}

We now explore the degree to which the superuniversality of Eq.~\eqref{2loopcorrelationexponent} persists away from this controlled large $N$ limit, i.e., when $k_{SU(N)}$ is continued back to its physical value given below Eq.~\eqref{simpleexplicit}.
We will use duality to argue that the physics described by ${\cal L}_m$ is independent of the particular value of $N$ appearing in the Lagrangian and that one consequence of this $N$ independence is that $\nu=1$ away from the controlled 't Hooft large $N$ limit.

In Secs.~\eqref{integersection} and \eqref{modularsection}, we showed that the effective Lagrangians describing the gapped phases that obtain from ${\cal L}_m$ for $M_\psi \neq 0$ do not depend on $N$.
It remains to argue that the physics of the intervening critical point could also be independent of $N$. 
For this, we conjecture a duality that equates the long wavelength behavior of the theory in \eqref{seed} to that of a free Dirac fermion for any integer $N$:
\begin{align}
\label{freediractogaugeddiracbulk}
i \bar{\Psi} \slashed{D}_A \Psi + {1 \over 2} {1 \over 4\pi} A d A  \longleftrightarrow {\cal L}_{\rm IQHT}(A). 
\end{align}
Remarkably, this duality implies that the physics described by ${\cal L}_{\rm IQHT}(A)$ does {\it not} depend on the particular value of $N$ appearing in its Lagrangian.
While a direct proof of Eq.~\eqref{freediractogaugeddiracbulk} is not known, we can show that Eq.~\eqref{freediractogaugeddiracbulk} is a consequence of the web of bosonization dualities in $2+1$ dimensions \cite{PhysRevX.6.031043, Seiberg:2016gmd} (see Appendices \ref{dualityargument} and \ref{particlehole} for details).
Furthermore, \eqref{freediractogaugeddiracbulk} is the statement of fermion particle-vortex duality \cite{Son2015, WangSenthilfirst2015, MetlitskiVishwanath2016, PhysRevX.6.031043,Seiberg:2016gmd} when $N=1$.
Consequently, the accumulated evidence for the duality web likewise provides support for Eq.~\eqref{freediractogaugeddiracbulk}.
In the remainder, we study the consequences of Eq.~\eqref{freediractogaugeddiracbulk}.

If the duality in Eq.~\eqref{freediractogaugeddiracbulk} holds for {\it all} integers $N \geq 1$, then $\nu$ must be independent of $N$ for the theory in \eqref{seed} and its ``modular descendants,", i.e., the theories of fractional quantum Hall transitions given by ${\cal L}_m$ in Eq.~\eqref{generallag}.
(See Appendix \ref{dualityargument} for the Abelian Chern-Simons dual of ${\cal L}_m$.)
Furthermore, choosing to determine $\nu$ at $N=\infty$, the specific value of $\nu$ should be captured at the leading planar order in the 't Hooft large $N$ limit.
This is because only planar terms scale as $N^0$ at large $N$; sub-planar terms always have an explicit dependence on $N$ through their $1/N$ prefactors (recall that both $k_{SU(N)}$ and $k_{U(1)}$ in Eq.~\eqref{simpleexplicit} scale linearly with $N$) and so they should not contribute to $\nu$ in any planar expansion at $N=\infty$.

Since $\nu = 1$ exactly for the theory of a free Dirac fermion, Eq.~\eqref{freediractogaugeddiracbulk} implies the planar contribution to $\gamma_{\bar{\psi} \psi}$ vanishes for the theory in \eqref{seed}. 
In the absence of non-perturbative corrections to the 't Hooft expansion in Eq.~\eqref{2loopcorrelationexponent} when $m \geq 2$, $\nu = 1$ should also hold for transitions involving fractional states, e.g., ${1 \over m +1} \rightarrow 0$ with $m \geq 2$, because $m$ only enters sub-planar terms in perturbation theory.
In other words, duality suggests the critical theories considered in this paper exhibit superuniversality with $\nu = z = 1$.

\section{Conclusion}

In this work, we introduced new effective theories with an emergent $U(N)$ gauge symmetry ($N > 1$) for various fractional quantum Hall transitions.
We showed that these theories are superuniversal in a controlled 't Hooft large $N$ limit and we argued that this conclusion holds more generality using duality.
Our theories function as an example that the effects of electron interactions and disorder can be disentangled from the phenomenon of superuniversality.
Furthermore, our theories provide examples of new dualities which are of fundamental interest and may have applications to other instances of quantum criticality.

There are several directions of further study.
It is important to better understand nonperturbative corrections to our theories; for instance, additional study of the lattice models in \cite{2012geraedtsmotrunich, 2016PhRvB..93c5103L} could provide useful insight. 
The theories in this paper may have direct application to quantum Hall transitions in graphene that can be controlled by varying an external electronic potential (\cite{2016arXiv161107113Z} and references therein).
Perhaps the most important direction is to incorporate the effects of disorder, which may account for the difference between the measured and theoretically determined correlation length exponent.

\begin{acknowledgments}
We thank S. Chakravarty, M. Fisher, E. Fradkin, T. Hartman, S. Kachru, S. Kivelson, O. Motrunich, D. Orgad, S. Raghu, S. Sachdev, E. Shimshoni, S. Sondhi, and S. Trivedi for helpful discussions and comments.
A.H. was supported by the National Science Foundation Graduate Research Fellowship under Grant No. DGE-1650441.
E.-A.K. was supported by the U.S. Department of Energy, Office of Basic Energy Sciences, Division of Materials Science and Engineering under Award de-sc0010313.
M.M. was supported in part by the UCR Academic Senate. 
M.M. is grateful for the hospitality of the Aspen Center for Physics, which is supported by the National Science Foundation (NSF) grant PHY-1607611.
The authors are grateful for the hospitality of the Kavli Institute for Theoretical Physics, under Grant No. NSF PHY-1125915.
\end{acknowledgments}

\onecolumngrid
\appendix
\section{Chern-Simons conventions}
\label{chernsimonsdefs}

In this appendix, we collect basic facts and definitions for Chern-Simons theories in $2+1$ dimensions.
The Chern-Simons term for the $U(N)$ gauge field $a$ is: 
\begin{align}
	{\rm Tr} \Big[a d a - {2 \over  3} i a^3 \Big] = N \epsilon^{\mu \nu \rho} (a^R_\mu \partial_\nu a^R_\rho - {2 \over 3} i f^{RST} a^R_\mu a^S_\nu a^T_\rho ),
\end{align}
where $a = a^R_\mu t^R$ for $U(N)$ (algebra) generators $t^R$ with $R \in \{1, \ldots, N^2\}$.
Our normalization convention for these generators is the following: ${\rm Tr}[t^R t^S] = N \delta^{R S}$ and $[t^R, t^S] = i f^{R S T} t^T$ where $f^{R ST}$ are the structure constants of $U(N)$.
We denote Abelian Chern-Simons terms:
\begin{align}
	A d B = \epsilon^{\mu \nu \rho} A_\mu \partial_\nu B_\rho,
\end{align}
where $\epsilon^{txy} = 1$.

In the absence of matter fields, only integral linear combinations of the following Chern-Simons terms appearing in Eq.~\eqref{seed} make well defined contributions to a $2+1$-dimensional effective action \cite{DeserWH, PolychronakosME}:
\begin{align}
	& {1 \over 4 \pi} {\rm Tr} \Big[a d a - {2 \over  3} i a^3 \Big], \cr
	& {1 \over 4 \pi} {\rm Tr} [a] d {\rm Tr} [a], \cr
	& {1 \over 2 \pi} {\rm Tr}[a] d b, \cr
	& {1 \over 4 \pi} b d b.
\end{align}
Since ${\rm Tr}[a]$ extracts the $U(1) \subset U(N)$ component of $a$, we can think of ${\rm Tr}[a]$ as a $U(1)$ gauge field with $2\pi$-quantized flux.
The combination of a single Dirac fermion and half-integer Chern-Simons level for $a$ in Eq.~\eqref{seed} yields a well defined term in the path integral \cite{NiemiRQ, RedlichDV, AlvarezGaumeWitten1984}.

We regularize our effective theories with a Yang-Mills term for $a$ and a Maxwell term for the Abelian gauge fields.
In a Yang-Mills regularization, the Chern-Simons level $k = - 1/2$ for the $SU(N) \subset U(N)$ component of $a$ receives a one-loop exact shift $k \rightarrow k + {\rm sign}(k) N$ \cite{witten1989, Chen:1992ee}. 
This correction arises from the interaction between the gauge fields contained in the Yang-Mills term.
If regularized by dimensional reduction \cite{Chen:1992ee}, the Chern-Simons level is not shifted (as the Yang-Mills interaction is no longer present).
To describe \eqref{seed} within dimensional reduction, the Chern-Simons level for the $SU(N)$ component $k_{\rm DR} = k + {\rm sign}(k) N$.

\section{Integer quantum Hall state and the insulator}
\label{integerphases}

In this appendix, we explain how the effective Lagrangian Eq.~\eqref{effectiveinteger} in the main text,
\begin{align}
	\label{effectiveintegerappendix}
	{\cal L}_{\rm eff}[A] & = {{\rm sign}(M_\psi) - 1 \over 2} {1 \over 4 \pi} {\rm Tr} \Big[a d a - {2 \over  3} i a^3 \Big] - {1 \over 2 \pi} {\rm Tr}[a] d b - {N + 1 \over 4 \pi} b d b - {1 \over 2 \pi} b d A,
\end{align}
describes an integer quantum Hall state when the fermion mass $M_\psi < 0$ and a topologically trivial insulator when $M_\psi > 0$.
In the effective Lagrangians written above and below, only relevant and marginal terms, in the renormalization group sense, are written; irrelevant operators (like Yang-Mills and Maxwell terms for the gauge fields) are understood to supplement ${\cal L}_{\rm eff}$ with a coefficient that scales inversely with the cutoff of the effective theory. 

Our strategy is to identify the integer quantum Hall state and the insulator through their electrical response to the $U(1)$, i.e., electromagnetic, gauge field $A$.
This response is encoded in an effective response Lagrangian, obtained by integrating out all dynamical degrees of freedom (e.g., $\psi$, $a$, and $b$ in Eq.~\eqref{seed}).
Consequently, this effective Lagrangian only contains $A$.
Using the relation $J_\mu = {\delta {\cal L}_{\rm eff}[A] \over \delta A^\mu}$, where $J_\mu$ is the electromagnetic current coupling to $A$, we can read off the electrical response to an applied electromagnetic field $A$. 
Focusing on the linear response of the system, we may terminate this effective Lagrangian at quadratic order in $A$.
As a simple example, consider the effective Lagrangian $\mathcal{L}_{CS} = \frac{1}{4\pi} AdA$ describing the integer quantum Hall state.
The relation, $J_i = \frac{1}{2\pi} \epsilon_{ij} E_j$, allows us to read off the Hall conductivity, $\sigma_{xy} = 1$, of this state, given in units where $e^2 = \hbar = 1$.

When $M_\psi < 0$, the effective Lagrangian takes the form:
\begin{align}
	\label{effectiveintegernegative}
	{\cal L}_{\rm eff}(M_\psi < 0) & = - {1 \over 4 \pi} {\rm Tr} \Big[a d a - {2 \over  3} i a^3 \Big] - {1 \over 2 \pi} {\rm Tr}[a] d b - {N + 1 \over 4 \pi} b d b - {1 \over 2 \pi} b d A.
\end{align}
We will show how Eq.~\eqref{effectiveintegernegative} describes an integer quantum Hall state by applying modular transformations to the rank/level duality $U(N)_1 \leftrightarrow SU(1)_N$ \cite{NACULICH1990687, Nakanishi1992, Hsin:2016blu}:
\begin{align}
	\label{simpleranklevel}
	-{1 \over 4 \pi} {\rm Tr} \Big[a d a - {2 \over  3} i a^3 \Big] - {1 \over 2 \pi} {\rm Tr}[a] d A \leftrightarrow {N \over 4 \pi}A d A.
\end{align}
Note that since $SU(1)$ is trivial, there are no dynamical gauge fields on the right-hand side. Eq.~\eqref{simpleranklevel} says that $U(N)$ Chern-Simons theory at level $k=-1$ is equivalent to the theory of $N$ copies of the $\nu=1$ integer quantum Hall state, i.e., a state with Hall conductivity equal to $N e^2/h$.
For instance, if the topological field theory on the left-hand side of the duality in \eqref{simpleranklevel} (or its dual on the right-hand side) is placed on a surface with boundary, there will be $N$ chiral Dirac fermions propagating along the edge.
We now sequentially act on both sides of the duality in \eqref{simpleranklevel} with ${\cal S} {\cal T}^{-N - 1}$, modular transformations defined in Eq.~\eqref{modulardef} in the main text.
First acting by ${\cal T}^{-N - 1}$, we obtain:
\begin{align}
	\label{ranklevelimplication}
	-{1 \over 4 \pi} {\rm Tr} \Big[a d a - {2 \over  3} i a^3 \Big] - {1 \over 2 \pi} {\rm Tr}[a] d A - {N +1 \over 4 \pi} A d A \leftrightarrow - {1 \over 4 \pi}A d A.
\end{align}
Then acting by ${\cal S}$, we find:
\begin{align}
	\label{ranklevelimplicationtwo}
	-{1 \over 4 \pi} {\rm Tr} \Big[a d a - {2 \over  3} i a^3 \Big] - {1 \over 2 \pi} {\rm Tr}[a] d b - {N +1 \over 4 \pi} b d b - {1 \over 2 \pi} b d A \leftrightarrow - {1 \over 4 \pi} c d c - {1 \over 2 \pi} c d A.
\end{align}
The theory on the left-hand side of the duality in \eqref{ranklevelimplicationtwo} is the effective Lagrangian ${\cal L}_{\rm eff}(M_\psi < 0)$ given in Eq.~\eqref{effectiveintegernegative}.
The theory on the right-hand side of \eqref{ranklevelimplicationtwo} is simply the effective hydrodynamic description of the integer quantum Hall effect \cite{wen1995}.
To see this, i.e., to see that the theory exhibits a Hall conductivity equal to one in units of $e^2/h$, we may integrate out $c$ using its equation of motion to find:
\begin{align}
	\label{ranklevelimplicationthree}
	- {1 \over 4 \pi} {\rm Tr} \Big[a d a - {2 \over  3} i a^3 \Big] - {1 \over 2 \pi} {\rm Tr}[a] d b - {N +1 \over 4 \pi} b d b - {1 \over 2 \pi} b d A \leftrightarrow {1 \over 4 \pi} A d A.
\end{align}

When $M_\psi > 0$, the effective Lagrangian,
\begin{align}
	\label{effectiveintegerpositive}
	{\cal L}_{\rm eff}(M_\psi > 0) & = - {1 \over 2 \pi} {\rm Tr}[a] d b - {N + 1 \over 4 \pi} b d b - {1 \over 2 \pi} b d A.
\end{align}
The $SU(N) \subset U(N)$ component of $a$ is no longer present in the effective Lagrangian.
Consequently, at low energies, it decouples from the remaining degrees of freedom: we assume that it confines.
The $U(1) \subset U(N)$ component of $a$, i.e., ${\rm Tr}[a]$, and $b$ remain in ${\cal L}_{\rm eff}(M_\psi > 0)$.
The equation of motion for ${\rm Tr}[a]$ sets $b=0$, up to gauge transformations.
Thus, 
\begin{align}
	{\cal L}_{\rm eff}(M_\psi > 0) = 0.
\end{align}
This Lagrangian describes a topologically trivial insulator as the Maxwell term for $A$ is understood to supplement ${\cal L}_{\rm eff}(M_\psi > 0)$.

A related way to see that ${\cal L}_{\rm eff}(M_\psi > 0)$ describes an insulator is to perform a $PSL(2, \mathbb{Z})$ field redefinition of the dynamical $U(1)$ gauge fields ${\rm Tr}[a] \mapsto \tilde{a}$ and $b \mapsto \tilde{b}$ so that ${\cal L}_{\rm eff}(M_\psi > 0) = {1 \over 4 \pi} \tilde{a} d \tilde{a} - {1 \over 4 \pi} \tilde{b} d \tilde{b} - {1 \over 2 \pi} (\tilde{a} - \tilde{b}) dA$ for odd $N$ or ${\cal L}_{\rm eff}(M_\psi > 0) = {1 \over 2\pi} \tilde{a} d \tilde{b} - {1 \over 2\pi} \tilde{a} d A$ for even $N$.
These effective Lagrangians describe topologically trivial insulators (if no symmetry is preserved) of fermions or bosons. 
There is no contradiction with the duality in \eqref{freediractogaugeddiracbulk} (or, alternatively, restriction to odd $N$), which says that Eq.~\eqref{seed} is dual to a free fermion, if we allow ourselves to ``stabilize" by a trivial insulator of fermions \cite{PhysRevB.89.115116}.

\section{Effective Lagrangians for fractional quantum Hall transitions}
\label{effectiveU1level}

In this appendix, we list the effective Lagrangians of the form given in Eq.~\eqref{generallag},
\begin{align}
	\label{generallagappendix1}
	{\cal L}_m & = {\cal L}_{\rm IQHT}(c) + {\cal L}_{\rm mod}(A),
\end{align}
where ${\cal L}_{\rm IQHT}(c)$ is given by Eq.~\eref{seed} with the replacement $A \rightarrow c$ and ${\cal L}_{\rm mod}(A)$ is determined by the particular modular transformation for a few other fractional quantum Hall transitions.
Because ${\cal L}_{\rm IQHT}(c)$ is the same in each effective Lagrangian, we only specify ${\cal L}_{\rm mod}(A)$.
We also determine the effective Chern-Simons level for $a_{U(1)}$ which scales as $|k_{U(1)}| \propto N$ for $N \rightarrow \infty$. 

\subsection{\texorpdfstring{$\sigma = 1/3 \rightarrow 2/5$}{1/3 -> 2/5} transition}

The $\sigma = 1/3 \rightarrow 2/5$ transition is obtained by acting on Eq.~\eqref{seed} by ${\cal S}^{-1} {\cal T}^{-2} {\cal S} {\cal T}$.
We find:
\begin{align}
	\label{simplefractionappnedix1}
	{\cal L}_{\rm mod}(A) = {1 \over 4 \pi} c d c - {1 \over 2 \pi} c d g - {2 \over 4 \pi} g d g + {1 \over 2 \pi} g d A.
\end{align} 
The corresponding effective Chern-Simons level for $a_{U(1)}$ in \eqref{simpleexplicit} for this transition is $\displaystyle k_{U(1)} = - {N \over 2} + {N^2 \over N + 5/3}$.

\subsection{\texorpdfstring{$\sigma = m/(m+1) \rightarrow 1$}{m/(m+1) -> 1} transition}

The $\sigma = m/(m+1) \rightarrow 1$ transition is obtained by acting on Eq.~\eqref{seed} by ${\cal S}^{-1} {\cal T}^{m} {\cal S} {\cal T}^{-1}$.
We find:
\begin{align}
	\label{simplefractionappendix2}
	{\cal L}_{\rm mod}(A) = - {1 \over 4 \pi} c d c - {1 \over 2 \pi} c d g + {m \over 4 \pi} g d g + {1 \over 2 \pi} g d A + {1 \over 4 \pi} A d A.
\end{align} 
The corresponding effective Chern-Simons level for $a_{U(1)}$ in \eqref{simpleexplicit} for this transition is $\displaystyle k_{U(1)} = - {N \over 2} + {N^2 \over N + 1/(m+1)}$.

\section{Fractional quantum Hall state and the insulator}
\label{appendixB}

In this appendix, we show how the effective Lagrangian in Eq.~\eqref{generallag} in the main text,
\begin{align}
	\label{generallagappendix2}
	{\cal L}_m & = {\cal L}_{\rm IQHT}(c) + {\cal L}_{\rm mod}(A),
\end{align}
where 
\begin{align}
	{\cal L}_{\rm IQHT}(c) & = i \bar{\psi} \not \! \! D_{a} \psi - {1 \over 2} {1 \over 4 \pi} {\rm Tr} \Big[a d a - {2 \over  3} i a^3 \Big] - {1 \over 2 \pi} {\rm Tr}[a] d b - {N + 1 \over 4 \pi} b d b - {1 \over 2 \pi} b d c
\end{align}
and
\begin{align}
	{\cal L}_{\rm mod}(A) = - {1 \over 2 \pi} c d g - {m \over 4 \pi} g d g + {1 \over 2 \pi} g d A,
\end{align}
describes a $1/(m+1) \rightarrow 0$ transition when $m \geq 0$.
Similar to Appendix \ref{integerphases}, when a fermion mass term $M_\psi \bar{\psi} \psi$ is added, we may integrate it out below the scale set by $|M_\psi|$ to find:
\begin{align}
	\label{effectivefractional}
	{\cal L}_{\rm eff} & = {{\rm sign}(M_\psi) - 1 \over 2} {1 \over 4 \pi} {\rm Tr} \Big[a d a - {2 \over  3} i a^3 \Big] - {1 \over 2 \pi} {\rm Tr}[a] d b - {N + 1 \over 4 \pi} b d b - {1 \over 2 \pi} b d c - {1 \over 2 \pi} c d g - {m \over 4 \pi} g d g + {1 \over 2 \pi} g d A.
\end{align}
We will show that Eq.~\eqref{effectivefractional} describes a fractional quantum Hall effect with Hall conductivity equal to $1/(m+1)$ (in units of $e^2/h$) when $M_\psi < 0$ and an insulator when $M_\psi > 0$. 

When $M_\psi < 0$, 
\begin{align}
	\label{effectivefractionalqh}
	{\cal L}_{\rm eff}(M_\psi < 0) = - {1 \over 4 \pi} {\rm Tr} \Big[a d a - {2 \over  3} i a^3 \Big] - {1 \over 2 \pi} {\rm Tr}[a] d b - {N + 1 \over 4 \pi} b d b - {1 \over 2 \pi} b d c - {1 \over 2 \pi} c d g - {m \over 4 \pi} g d g + {1 \over 2 \pi} g d A.
\end{align}
Applying ${\cal S}^{-1} {\cal T}^{-m} {\cal S}^2 {\cal T}^{-N - 1}$ to the rank/level dual pair \cite{NACULICH1990687, Nakanishi1992, Hsin:2016blu} in \eqref{simpleranklevel}, we find:
\begin{gather}
	\label{anotherimplication}
	- {1 \over 4 \pi} {\rm Tr} \Big[a d a - {2 \over  3} i a^3 \Big] - {1 \over 2 \pi} {\rm Tr}[a] d b - {N + 1 \over 4 \pi} b d b - {1 \over 2 \pi} b d c - {1 \over 2 \pi} c d g - {m \over 4 \pi} g d g + {1 \over 2 \pi} g d A \cr
	\updownarrow \cr 
	- {1 \over 4 \pi} b d b - {1 \over 2 \pi} b d c - {1 \over 2 \pi} c d g - {m \over 4 \pi} g d g + {1 \over 2 \pi} g d A.
\end{gather}
Thus, ${\cal L}_{\psi}(M_\psi < 0)$ (the theory in the top line of \eqref{anotherimplication}) is dual to the theory in the bottom line of \eqref{anotherimplication}.
We now sequentially integrate out $b$ and $c$ so that the bottom line of \eqref{anotherimplication} simplifies to
\begin{align}
	- {m + 1 \over 4 \pi} g d g + {1 \over 2 \pi} g d A.
\end{align}
This is the hydrodynamic effective Lagrangian for the fractional quantum Hall state with Hall conductivity equal to $1/(m+1)$ \cite{wen1995}.
Thus, we find:
\begin{align}
	- {1 \over 4 \pi} {\rm Tr} \Big[a d a - {2 \over  3} i a^3 \Big] - {1 \over 2 \pi} {\rm Tr}[a] d b - {N + 1 \over 4 \pi} b d b - {1 \over 2 \pi} b d c - {1 \over 2 \pi} c d g - {m \over 4 \pi} g d g + {1 \over 2 \pi} g d A \leftrightarrow - {m + 1 \over 4 \pi} g d g + {1 \over 2 \pi} g d A.
\end{align}

When $M_\psi > 0$, 
\begin{align}
	\label{effectivefractionalins}
	{\cal L}_{\rm eff}(M_\psi > 0) = - {1 \over 2 \pi} {\rm Tr}[a] d b - {N + 1 \over 4 \pi} b d b - {1 \over 2 \pi} b d c - {1 \over 2 \pi} c d g - {m \over 4 \pi} g d g + {1 \over 2 \pi} g d A. 
\end{align}
The $SU(N) \subset U(N)$ component of $a$ again decouples and we assume it confines.
The equation of motion for ${\rm Tr}[a]$ sets $b = 0$; the equation of motion for $c$ sets $g = 0$ and we are left with the effective Lagrangian for an insulator:
\begin{align}
	{\cal L}_{\rm eff}(M_\psi > 0) = 0.
\end{align}

\section{Duality argument and Abelian Chern-Simons duals}
\label{dualityargument}

\subsection{Duality argument}

In the first part of this appendix, we argue that Eq.~\eqref{seed} is in the same universality class as a free fermion.
Our demonstration applies the argument of \cite{PhysRevX.6.031043, Seiberg:2016gmd} to the bosonization duality \cite{GMPTWY2012, AharonyGurAriYacoby2012, AharonyGurAriYacobysecond2012, Aharony2016,Hsin:2016blu},
\begin{gather}
	\label{wfgaugeddirac}
	|D_A \phi|^2 - |\phi|^4 + {1 \over 4 \pi} A d A \cr
	\updownarrow \\
	i \bar{\psi} D_a \psi - {1\over 8\pi} {\rm Tr}[a d a - {2 \over 3} i a^3] - {1 \over 2\pi} {\rm Tr}[a] d A - {N-1 \over 4 \pi} A d A, \nonumber
\end{gather}
that relates the theory of a Wilson-Fisher boson $\phi$ to the theory of a $U(N)$ Chern-Simons gauge field $a$ coupled to a Dirac fermion $\psi$.
Applying the modular transformation ${\cal S} {\cal T}^{-2}$ to ``both sides" of this duality (we introduce $c$ in the Wilson-Fisher theory and $b$ in the gauged Dirac theory in applying the ${\cal S}$ transformation), we find the low-energy equivalence:
\begin{align}
	\label{gaugedwfmodifiedgaugeddirac}
	|D_c \phi|^2 - |\phi|^4 - {1 \over 4 \pi} c d c - {1 \over 2 \pi} c d A \leftrightarrow
	{\cal L}_{\rm IQHT}(A),
\end{align}
with ${\cal L}_{\rm IQHT}(A)$ given in Eq.~\eqref{seed}.
But the gauged Wilson-Fisher theory on the left-hand side of \eqref{gaugedwfmodifiedgaugeddirac} is also dual to the theory of a free Dirac fermion \cite{PhysRevD.94.085009,PhysRevLett.118.011602, 2017arXiv170505841C, 2017arXiv170501106M}. 
Thus, we relate the low-energy physics of the theory of a free Dirac fermion to that of our theory in Eq.~\eqref{seed},
\begin{align}
	\label{freediractogaugeddirac}
	i \bar{\Psi} \slashed{D}_A \Psi + {1 \over 2} {1 \over 4\pi} A d A  \leftrightarrow {\cal L}_{\rm IQHT}(A). 
\end{align}

\subsection{Abelian Chern-Simons duals}

In the second part of this appendix, we provide the Abelian Chern-Simons duals for the $U(N)$ Chern-Simons theories studied in the main text and listed in Appendix \ref{effectiveU1level} that are implied by the duality in \eqref{freediractogaugeddiracbulk} (copied below):
\begin{align}
	\label{freediractogaugeddiracappendix}
	i \bar{\Psi} \slashed{D}_A \Psi + {1 \over 2} {1 \over 4\pi} A d A  \leftrightarrow {\cal L}_{\rm IQHT}(A). 
\end{align}
The strategy is identical to that of \cite{Seiberg:2016gmd}: we perform a modular transformation on each side of the duality \eqref{freediractogaugeddiracappendix} and then identify the resulting theories.
Duality implies that 't Hooft large $N$ limit calculations for the theories with non-Abelian gauge group can be reinterpreted in terms of their Abelian duals.

\subsection{Dual pair for the \texorpdfstring{$\sigma = 1/(m+1) \rightarrow 0$}{1/m -> 0} transition}

Acting on \eqref{freediractogaugeddiracappendix} with ${\cal S}^{-1} {\cal T}^{-m} {\cal S}$, we find the duality:
\begin{gather}
	i \bar{\Psi} \slashed{D}_{\tilde{a}} \Psi + {1 \over 2} {1 \over 4\pi} \tilde{a} d \tilde{a} - {1 \over 2\pi} \tilde{a} d \tilde{b} - {m \over 4 \pi} \tilde{b} d \tilde{b} + {1 \over 2 \pi} \tilde{b} d A \cr
	\updownarrow \\
	i \bar{\psi} \not \! \! D_{a} \psi - {1 \over 2} {1 \over 4 \pi} {\rm Tr} \Big[a d a - {2 \over  3} i a^3 \Big] 
	- {1 \over 2 \pi} {\rm Tr}[a] d b - {N + 1 \over 4 \pi} b d b - {1 \over 2 \pi} b d c - {1 \over 2 \pi} c d g - {m \over 4 \pi} g d g + {1 \over 2 \pi} g d A \nonumber
\end{gather}
where $\tilde{a}$, $\tilde{b}$, $b$, $c$, and $g$ are Abelian gauge fields and $a$ is a $U(N)$ gauge field.

\subsection{Dual pair for the \texorpdfstring{$\sigma = 1/3 \rightarrow 2/5$}{1/3 -> 2/5} transition}

Acting on \eqref{freediractogaugeddiracappendix} with ${\cal S}^{-1} {\cal T}^{-2} {\cal S} {\cal T}$, we find the duality:
\begin{gather}
	i \bar{\Psi} \slashed{D}_{\tilde{a}} \Psi + {3 \over 2} {1 \over 4\pi} \tilde{a} d \tilde{a} - {1 \over 2\pi} \tilde{a} d \tilde{b} - {2 \over 4 \pi} \tilde{b} d \tilde{b} + {1 \over 2 \pi} \tilde{b} d A \cr
	\updownarrow \\
	i \bar{\psi} \not \! \! D_{a} \psi - {1 \over 2} {1 \over 4 \pi} {\rm Tr} \Big[a d a - {2 \over  3} i a^3 \Big] 
	- {1 \over 2 \pi} {\rm Tr}[a] d b - {N + 1 \over 4 \pi} b d b - {1 \over 2 \pi} b d c + {1 \over 4 \pi} c d c - {1 \over 2 \pi} c d g - {2 \over 4 \pi} g d g + {1 \over 2 \pi} g d A \nonumber
\end{gather}
where $\tilde{a}$, $\tilde{b}$, $b$, $c$, and $g$ are Abelian gauge fields and $a$ is a $U(N)$ gauge field.

\subsection{Dual pair for the \texorpdfstring{$\sigma = m/(m+1) \rightarrow 1$}{m/(m+1) -> 1} transition}

Acting on \eqref{freediractogaugeddiracappendix} with ${\cal T} {\cal S}^{-1} {\cal T}^{m} {\cal S} {\cal T}^{-1}$, we find the duality:
\begin{gather}
	i \bar{\Psi} \slashed{D}_{\tilde{a}} \Psi - {1 \over 2} {1 \over 4\pi} \tilde{a} d \tilde{a} - {1 \over 2\pi} \tilde{a} d \tilde{b} + {m \over 4 \pi} \tilde{b} d \tilde{b} + {1 \over 2 \pi} \tilde{b} d A + {1 \over 4\pi} A d A\cr
	\updownarrow \\
	i \bar{\psi} \not \! \! D_{a} \psi - {1 \over 2} {1 \over 4 \pi} {\rm Tr} \Big[a d a - {2 \over  3} i a^3 \Big] 
	- {1 \over 2 \pi} {\rm Tr}[a] d b - {N + 1 \over 4 \pi} b d b - {1 \over 2 \pi} b d c - {1 \over 4 \pi} c d c - {1 \over 2 \pi} c d g + {m \over 4 \pi} g d g + {1 \over 2 \pi} g d A + {1 \over 4 \pi} A d A \nonumber
\end{gather}
where $\tilde{a}$, $\tilde{b}$, $b$, $c$, and $g$ are Abelian gauge fields and $a$ is a $U(N)$ gauge field.

\section{Particle-hole transformation within the lowest Landau level}
\label{particlehole}

For the free Dirac theory in the duality in \eqref{freediractogaugeddiracbulk}, the particle-hole transformation with respect to a filled Landau level can be defined as follows.
First, the fields are transformed by the anti-unitary ($i \mapsto - i$) transformation that consists of the product of time-reversal and charge-conjugation which takes $t \mapsto - t$,
\begin{align}
	\Psi & \mapsto - \gamma^t \Psi^\ast, \cr
	(A_t, A_x, A_y) & \mapsto (- A_t, A_x, A_y),
\end{align}
and then the Lagrangian is shifted by a filled Landau level using the ${\cal T}$ transformation.

The theory of a free Dirac fermion in \eqref{freediractogaugeddiracbulk} is invariant under a particle-hole transformation with respect to a filled Landau level.
Duality implies that the theory in Eq.~\eqref{seed} likewise enjoys this symmetry; we believe particle-hole symmetry is realized quantum mechanically and is not visible in the classical Lagrangian of Eq.~\eqref{seed} for $N > 1$ (see \cite{Aharony2017} for a recent discussion of this phenomena in related dualities).
It would be interesting to see how this symmetry constrains the conductivity (along with other observables) of different quantum critical states \cite{PhysRevB.57.7157, 2016JHEP...07..090G}.

There is second anti-unitary transformation that we expect to leave physical observables invariant even though it is not a symmetry of Eq.~\eqref{seed}.
It is defined as follows: first, time-reversal acts on the dynamical fields as
\begin{align}
	\psi & \mapsto \gamma^y \psi, \cr
	(a_t, a_x, a_y) & \mapsto (a_t, - a_x, - a_y), \cr
	(b_t, b_x, b_y) & \mapsto (b_t, - b_x, -b_y);
\end{align} 
second, the product of time-reversal and charge-conjugation acts on $A$ as
\begin{align}
	(A_t, A_x, A_y) \mapsto (- A_t, A_x, A_y);
\end{align}
Finally, the Lagrangian in Eq.~\eqref{seed} is shifted by a filled Landau level with the ${\cal T}$ transformation. 
This transformation can be employed to generate alternative effective descriptions for the particle-hole conjugate of a given quantum Hall phase transition.

\twocolumngrid
\bibliography{bigbib}
\end{document}